\UseRawInputEncoding

\documentclass[11pt,tightenlines,aps,prd,groupedaddress,nofootinbib,nobibnotes,longbibliography,notitlepage]{revtex4-1}
\pdfoutput=1

\usepackage{graphicx}
\usepackage{amsfonts}
\usepackage{amssymb}
\usepackage{amsbsy}
\usepackage{amsmath}
\usepackage{mathtools}
\usepackage{latexsym}
\usepackage{bm}
\usepackage{color}
\usepackage{hyperref}
\usepackage[margin=2cm]{geometry}
\usepackage{comment}
\usepackage[makeroom]{cancel}
\usepackage[dvipsnames]{xcolor}

\usepackage{stackengine}
\stackMath
\newcommand\tenq[2][1]{%
 \def\useanchorwidth{T}%
  \ifnum#1>1%
    \stackon[0pt]{\tenq[\numexpr#1-1\relax]{#2}}{\scriptscriptstyle\sim}%
  \else%
    \stackon[1pt]{#2}{\scriptscriptstyle\sim}%
  \fi%
}

\def\nn {\nonumber}

\newcommand{\de}{\mbox{d}}
\newcommand{\lf}{\left}
\newcommand{\rg}{\right}
\newcommand{\be}{\begin{equation}}
\newcommand{\ee}{\end{equation}}

\newcommand{\pa}{\partial}

\newcommand{\bea}{\begin{eqnarray}}
\newcommand{\eea}{\end{eqnarray}}
\newcommand{\scr}{\scriptscriptstyle}

\numberwithin{equation}{section}

\renewcommand{\theequation}{\arabic{section}.\arabic{equation}}

\begin{document}

\title{\Large Evolving black hole with scalar field accretion} 

\author{Marco de Cesare}
\email{marco.decesare@ehu.eus}

\affiliation{Department of Physics, University of the Basque Country UPV/EHU, 48940 Leioa, Spain}

\author{Roberto Oliveri}
\email{roberto.oliveri@obspm.fr}

\affiliation{LUTH, Laboratoire Univers et Th\'eories, Observatoire de Paris, CNRS, Universit\'e PSL, Universit\'e Paris Cit\'e, 5 place Jules Janssen, 92190 Meudon, France}

%\date{\today}

\begin{abstract}
We obtain approximate analytical solutions of the Einstein equations close to the trapping horizon for a dynamical spherically symmetric black hole in the presence of a minimally coupled self-interacting scalar field. This is made possible by a new parametrization of the metric, in which the displacement from the horizon as well as its expansion rate feature explicitly. Our results are valid in a neighbourhood of the horizon and hold for any scalar field potential and spacetime asymptotics.
An exact equation for the accretion rate is also obtained, which generalizes the standard Bondi formula. 
We also develop a dynamical system approach to study near-equilibrium black holes; using this formalism, we focus on a simple model to show that the near-equilibrium dynamics is characterised by scaling relations among dynamical variables.
Moreover, we show that solutions with purely ingoing energy-momentum flux never reach equilibrium.

\end{abstract}

\maketitle

\tableofcontents

\nopagebreak

\newpage

\section{Introduction}
In an expanding universe the accretion of a black hole is determined by its interactions with the cosmological fluid medium, whose equation of state and energy density are in general time dependent. In general relativity the dynamics of black hole observables, matter fields, and the cosmological scale factor are non-linearly coupled: even assuming spherical symmetry, finding a solution of the equations of motion is a highly challenging task. Some earlier attempts are reviewed in Ref.~\cite{Carrera:2008pi}. Standard Newtonian approximations for spherical accretion \cite{Bondi} and the general relativistic Michel solution \cite{Michel:1972wc} both assume a stationary fluid flow and do not take into account cosmological expansion; they are therefore inadequate to study the evolution of black holes over cosmological time scales. Moreover, there are no known exact solutions of the Einstein equations that describe physically realistic evolving black holes with matter. The well-known McVittie solution \cite{McVittie,Nolan:1998xs,Nolan:1999kk} has several shortcomings: it describes a non-accreting black hole and, beyond the special case of Schwarzschild-de Sitter, it features a spatially homogeneous energy density while the pressure is inhomogeneous \cite{Kaloper:2010ec}. Another possibility that is often considered is the embedding of a Schwarzschild black hole into an otherwise homogeneous and isotropic Friedmann-Lema{\^i}tre-Robertson-Walker universe by imposing suitable junction conditions; this is known as the Einstein-Straus model \cite{Einstein:1945id,Mars:2013ooa}. The main drawback of the Einstein-Straus model is that it is consistent only if the cosmological background is filled with pressureless dust, otherwise the junction is not smooth at the matching surface \cite{Mars:2013ooa}.
In the case of a massless scalar field an exact non-static solution was obtained in Ref.~\cite{Husain:1994uj} that describes a black hole in an expanding cosmological background; however, this solution has a timelike naked singularity and it is not known how to generalize it to include a potential for the scalar field.\footnote{See also Ref.~\cite{Husain:1995bf} for exact non-static black hole solutions sourced by a null fluid.}
All these reasons motivate us to develop a new approach to study evolving black holes in cosmology using analytical techniques.
To illustrate our approach in the simplest yet non-trivial case, we consider a self-interacting real scalar field as matter. Scalar fields play an important role in cosmology, especially in the early universe where they drive the accelerated expansion during inflation.
In this paper, we study the evolution of a spherically symmetric black hole in the presence of a minimally coupled self-interacting scalar field.
In particular, we find an exact accretion law for the apparent horizon (more precisely, the {\it future outer trapping horizon} \cite{Hayward:1993wb}).
By suitably matching the near-horizon asymptotics to the cosmological evolution, our results may find direct application to the evolution of black holes during inflation ---without imposing any simplifying assumptions, such as slow roll, which was used in previous studies \cite{Chadburn:2013mta,Gregory:2018ghc}.
Our approach relies on a systematic expansion of the Einstein field equations around the black-hole apparent horizon in terms of a (compact) near-horizon coordinate $z$. The field equations are then solved order by order in $z$.
In the new coordinates, the metric explicitly depends also on the horizon expansion rate $\dot{r}_H$; this enables us to study different dynamical regimes of the system analytically. From the first-order field equations we obtain an exact accretion equations that generalizes the standard Bondi formula.
Moreover, as an interesting example, we specialize to Neumann boundary conditions for the scalar field at the horizon, which single out a unique solution characterized by purely ingoing energy-momentum flux at the horizon
\footnote{A purely ingoing flux at the apparent horizon $T_{ab}n^a n^b=0$ has been considered earlier as a boundary condition for a null fluid with traceless stress-energy tensor in Ref.~\cite{Gunasekaran:2019jrq}}, and we explicitly solve the equations to third order in $z$. We show that Neumann boundary conditions are incompatible with the static limit for this system.
In addition, we study the approach to equilibrium by simultaneously expanding the field equations around a static Schwarzschild-de~Sitter solution and around the horizon. This enables us to map the gravitational field equations to an infinite dimensional dynamical system; explicit solutions are then obtained for a truncation of this system. Furthermore, assuming a simple model for the decay of the scalar field at the horizon, we show that the approach to equilibrium is characterised by scaling relations, independent of the functional form of the potential or any parameters of the model.
We also obtain a {\it first law} for the evolving black hole with scalar field and discuss the thermodynamical interpretation for this physical system.
The paper is organized as follows. In Section~\ref{Sec:2} we introduce a new coordinate system adapted to the evolving horizon, derive the equations of motion and present our solution scheme. In Section~\ref{Sec:Geo} we review basic geometric properties of the horizon and compute the energy-momentum fluxes in the ingoing and outgoing null directions; furthermore, we derive exact Bondi-like accretion laws for the horizon. In Section~\ref{Sec:Neumann} we impose Neumann boundary conditions for the scalar field at the horizon and derive the corresponding solution; moreover, we show that this solution cannot approach equilibrium.
In Section~\ref{Sec:Adiabatic} we expand the field equations around the static Schwarzschild-de~Sitter solution and obtain a dynamical system description of the dynamics; particular attention is paid to the approach to equilibrium. In Section~\ref{Sec:Thermodynamics} we derive the first law of black hole dynamics. Finally, in Section~\ref{Sec:Discussion} we review our results and discuss prospects for future work.

%%%%%%%%%%%%%%%%%
%SECTION 2
%%%%%%%%%%%%%%%%%

\section{Near-horizon dynamics} \label{Sec:2}
To study the dynamical evolution of a black hole, it is convenient to choose a coordinate system that is regular in a neighbourhood of the horizon. In Eddington-Finkelstein coordinates the metric of a general spherically symmetric geometry can be expressed as
\be\label{Eq:Metric}
\de s^2= - e^{2\beta(v,r)}A(v,r) \de v^2 +2 e^{\beta(v,r)}\de v \de r +r^2 \de \Omega^2 ~,
\ee
where $r$ is the areal radius and $v$ is constant along ingoing radial null geodesics. The metric function $A(v,r)$ is parametrized as
\be
A(v,r)=1- \frac{2G\, m(v,r)}{r} ~,
\ee
where $m(v,r)$ is the Misner-Sharp mass, measuring the mass contained within spheres of radius $r$. The position of the apparent horizon, defined as a marginally trapped surface, is determined by the zeroes of $A(v,r)$ \cite{Bengtsson:2010tj}
\be\label{Eq:HorizonDef}
r=2G\, m(v,r)~.
\ee
In turn, Eq.~\eqref{Eq:HorizonDef} implicitly defines the horizon as a function of $v$, $r_H=r_H(v)$. In general Eq.~\eqref{Eq:HorizonDef} admits multiple solutions, corresponding, \emph{e.g.}, to a cosmological horizon in addition to the black hole horizon. Thus, in case there are multiple horizons, in order to identify a solution of \eqref{Eq:HorizonDef} as the black hole horizon,\footnote{More precisely a {\it future outer trapping horizon}, that is foliated by marginally trapped spheres with areal radius $r_H(v)$.} we shall also demand that the following conditions be satisfied: $\theta_n<0$, ${\cal L}_n \theta_l<0$ \cite{Hayward:1993wb,Booth:2005qc}.
 Here $l^a$ and $n^a$ denote, respectively, the outgoing and ingoing radial null vectors, $\theta_l$ is the expansion of $l^a$ and $\theta_l=0$ at the horizon.
We introduce a near-horizon coordinate $z$, defined through
\be
r(v,z)=\frac{r_H(v)}{1-z} ~.
\ee
Compared to other possible choices of near-horizon coordinates it has the advantage of being compact, which is convenient in numerical resolution approaches. In the static case, a similar coordinate has been introduced earlier in Ref.~\cite{Rezzolla:2014mua}.
The apparent horizon is now located at $z=0$, whereas spatial infinity is at $z=1$.\footnote{Note that there may be coordinate singularities in the interval $0< z < 1$ if there is more than one horizon. For instance, in Schwarzschild-de Sitter there is a cosmological horizon located at $z=1- r_H H$, where $H$ is the (constant) Hubble rate.}
 In terms of the new coordinate $z$, the function $A(v,z)$ has a zero at $z=0$ and admits the following expansion
\be\label{Eq:z_expansion_A}
A(v,z)= \left( 1 - \frac{2 G\pa_z m \vert_{z=0}}{r_H}\right)\, z +\frac{G\left(2 \pa_z m - \pa_z^2 m\right)\vert_{z=0}}{r_H}\, z^2 +  \dots ~.
\ee
The extremal case arises when the first-order term vanishes in Eq.~\eqref{Eq:z_expansion_A}, whereby $A$ has a higher-order zero at the horizon. In this paper we will focus specifically on non-extremal black holes, although our method can be generalized to the extremal case as well. When the metric is expressed in $(v,z)$ coordinates, the horizon expansion rate $\dot{r}_H$ also appears explicitly; this enables us to have more direct control on the different dynamical regimes of the system, see Section~\ref{Sec:Adiabatic}
\be
\de s^{2} = \left(-e^{2\beta(v,z)}A(v,z) + \frac{2e^{\beta(v,z)}\dot{r}_{H}(v)}{1-z}\right)\de v^{2} +\frac{2e^{\beta(v,z)} r_{H}(v) }{(1-z)^{2}}\de v \de z + \frac{r_{H}^{2}(v)}{(1-z)^{2}}\de \Omega^2 ~.
\ee

The Einstein field equations, assuming a minimally-coupled self-interacting scalar field as matter source, read as (here $\kappa = 8\pi G$)
\be\label{Eq:EinsteinFieldEquations}
G_{ab}=\kappa\, T_{ab}=\kappa \lf[\pa_a\phi\pa_b\phi -g_{ab}\lf( \frac{1}{2}g^{cd}\pa_c \phi \pa_d \phi +U(\phi)\rg) \rg] ~.
\ee
A cosmological constant term can be reabsorbed in the definition of the scalar potential.\footnote{If required, the cosmological constant can be reinstated explicitly by means of the shift $U(\phi)\rightarrow U(\phi) + \kappa^{-1}\Lambda$.}
In $(v,z)$ coordinates, we obtain after some rearrangements\footnote{Hereinafter, derivatives with respect to~$z$ are denoted by a prime and derivatives with respect to~$v$ are denoted by a dot.}
\begin{subequations} \label{Eq:FieldEquations}
\begin{align}
&2 \beta' = \kappa(1-z) \lf(\phi' \rg)^2~,\label{EQ:efe0i}\\
&1 - A - (1-z) \left(A' +A\beta'\right)= \frac{ \kappa r_H^2}{(1-z)^2} U(\phi)~,\label{EQ:efe00}\\
&\frac{\dot{A}}{1-z} - \frac{\dot{r}_H }{r_H} A'  =\kappa\left(\frac{ \dot{r}_H}{r_H} \phi' - 
   \frac{\dot{\phi}}{1-z} \right)\lf[ (1-z) \phi'  A-e^{-\beta} r_H \left( \frac{\dot{r}_H}{r_H} \phi' - \frac{\dot{\phi}}{1-z}\right)\rg] ~, \label{EQ:efei0}\\
&2\beta'' A + A'' + \beta' \left( 3A' +2 \beta'  A\right) - \frac{2e^{-\beta}r_{H}}{1-z}\left(\frac{\dot{r}_{H}}{r_{H}}\beta'' - \frac{\dot{\beta}'}{1-z}\right)
= \frac{\kappa e^{-\beta }\,r_{H}\phi' }{1-z}\left( \frac{\dot{r}_{H}}{r_{H}}\phi'  - \frac{2\, \dot{\phi} }{1-z}\right) -  \frac{2\kappa\,  r_H^2 U(\phi )}{(1-z)^4} ~.\label{EQ:efetheta}
\end{align}
\end{subequations}
The Klein-Gordon equation follows from Eq.~\eqref{Eq:EinsteinFieldEquations} using the contracted Bianchi identities $\nabla^a G_{ab}=0$. It reads as
\be\label{EQ:KGequation}
\begin{split}
\dot{\phi}'+  \frac{\dot{\phi}}{1-z} -\frac{\dot{r}_{H}}{r_{H}}(1-z)\phi'' +\frac{e^{\beta}(1-z)^{2}}{2r_{H}}\left(A\phi'' +A' \phi' + A \phi'\beta' \right)-\frac{e^\beta r_H}{2(1-z)^2}   \frac{\pa U}{\pa \phi}=0~.
\end{split}
\ee
We require that both the geometry and matter be regular at the horizon. Therefore, we assume that the unknown functions $A$, $\beta$ and $\phi$ admit a power-series expansion in $z$
\be\label{Eq:Expansions}
A(v,z)=\sum_{n=1}^{\infty} a_n(v) z^n ~, \quad
\beta(v,z)= \sum_{n=1}^{\infty} b_n(v) z^n ~, \quad
\phi(v,z)= \phi_o(v) \left(1+\sum_{n=1}^{\infty} c_n(v) z^n \right) ~.
\ee
Note that we can set $b_0(v)=0$ without loss of generality, since this amounts to a $z$-independent redefinition of the $v$ coordinate.
For a non-extremal black hole we have $a_1\neq0$. In particular, we shall assume that $a_1>0$; this condition ensures that the horizon is outer trapping, {\it i.e.}~${\cal L}_n \theta_l\lvert_{z=0}<0$, see Eqs.~\eqref{Eq:OuterTrap}

Substituting the power-series expansions \eqref{Eq:Expansions} into the equations of motion, we find that the first-order coefficients must satisfy\footnote{We omit the time dependence to make the notation lighter.}
\be\label{Eq:FirstOrder}
a_1=1-\kappa\, r_H^2 U(\phi_o)>0 ~,~~
b_1=\frac{1}{2}\kappa c_1^2 \phi_o^2~,~~
a_1 \dot{r}_H=\kappa  \left(c_1 \phi_o \dot{r}_H - r_H \dot{\phi}_o\right)^2 ~,
\ee
Equations for the higher order coefficients can similarly be obtained but are omitted for brevity. By comparison with the $z$-expansion of the Schwarzschild-de~Sitter solution, the solution for $a_1$ in Eq.~\eqref{Eq:FirstOrder} shows that the quantity $\kappa\, U(\phi_o)$ plays the role of an effective (time-dependent) cosmological constant. Nevertheless, for generic solutions the correspondence with the Schwarzschild-de~Sitter spacetime is broken by higher-order corrections and therefore only holds in the close proximity of the horizon.

\section{Geometry and matter in the near-horizon region: accretion law}\label{Sec:Geo}
Before examining the solutions of the field equations in the following sections, we recall some basic properties of the apparent horizon and compute the energy-momentum fluxes along null directions. These properties are useful in order to physically interpret the solutions and their boundary conditions; in particular, they enable us to extract an accretion law for the black hole solely from the first-order solutions \eqref{Eq:FirstOrder}.

To study the spacetime geometry in the near-horizon region we introduce outgoing and ingoing radial null vectors, denoted respectively as $l^a$ and $n^a$. These are given by (see Ref.~\cite{Bengtsson:2010tj}, although note the different notation)
\be
 l^a=\left(\frac{\pa}{\pa v}\right)^a+\frac{e^{\beta}}{2}A\left(\frac{\pa}{\pa r}\right)^a ~, \quad  n^a=-e^{-\beta}\left(\frac{\pa}{\pa r}\right)^a ~,
\ee
with the normalization $l^a n_a=-1$. The corresponding expansion scalars are
\be
 \theta_l=\frac{e^\beta}{r}A=  \frac{a_1}{r_H} z + \mathcal{O}(z^2) ~,\quad \theta_n= -\frac{2e^{-\beta}}{r} = -\frac{2}{r_H}  +\mathcal{O}(z) ~.
\ee
The Lie derivative of the expansion scalar associated with $l^{a}$ with respect to the ingoing radial vector $n^{a}$ is given by
\be\label{Eq:OuterTrap}
{\cal L}_n  \theta_l =  \left(1-r\pa_r \beta\right)\frac{A}{r^2} -\frac{\pa_r A}{r} = -\frac{a_1}{r_H^2} + \mathcal{O}(z)~.
\ee
Since $a_1>0$, this shows that the $z=0$ surface is indeed a {\it future outer trapping horizon} according to Hayward's definition~\cite{Hayward:1993wb}. The positivity of $a_1$ also implies that the marginally trapped tube $r=r_H(v)$ is spacelike, see Ref.~\cite{Booth:2005ng} [Eq.~(2.13) therein].
In the null basis and to lowest order in $z$, the stress-energy tensor has components
\be\label{Eq:StressEnergyNullComponents}
T_{ab}n^a n^b\Big\lvert_{z=0}=  \frac{c_1^2\, \phi_o^2}{r_H^2}~,~~T_{ab}l^a l^b\Big\lvert_{z=0}= \lf(\dot{\phi}_o-c_1\phi_o \frac{\dot{r}_H}{r_H}\rg)^2 ~, ~~
T_{ab}n^a l^b\Big\lvert_{z=0} = U(\phi_o)~.
\ee
The component $T_{ab}n^a n^b\lvert_{z=0}$ represents the flux of energy-momentum along the outgoing null direction $l^a$ and vanishes if and only if $c_1=0$. Combining Eqs.~\eqref{Eq:FirstOrder} and \eqref{Eq:StressEnergyNullComponents} we obtain the accretion law
\be\label{Eq:AccretionGeneral}
\dot{r}_H=\frac{\kappa\, r_{H}^2}{1-{\cal E}}T_{ab}l^a l^b=\frac{\kappa\, r_{H}^2 }{1-{\cal E}}({\cal L}_l \phi\lvert_{z=0})^2 \geq 0~,
\ee
where we defined $\mathcal{E}\equiv\kappa\, r_H^2 U(\phi_o)$ and in the last step we used ${\cal L}_l \phi\lvert_{z=0}=\dot{\phi}_o-c_1\phi_o \frac{\dot{r}_H}{r_H}$~. It follows that the black hole horizon is a monotonically increasing function of $v$. This is in agreement with the area increase law~\cite{Ashtekar:2004cn} since the scalar field satisfies the null energy condition.
The accretion law \eqref{Eq:AccretionGeneral} can also be derived geometrically using the results in Ref.~\cite{Booth:2005ng} (Eq.~(2.13) therein).\footnote{We thank Ivan Booth for pointing out the connection with Ref.~\cite{Booth:2005ng}.}

Lastly, we can rewrite \eqref{Eq:AccretionGeneral} in terms of the black hole mass $M(v)\equiv m(v,r_H)=r_H(v)/(2G)$
\be\label{Eq:AccretionGeneralMass}
\dot{M}=\frac{16 \pi  G^2}{1-\mathcal{E}}  M^2 ({\cal L}_l \phi\lvert_{z=0})^2~.
\ee
The accretion law~\eqref{Eq:AccretionGeneralMass} is exact and does not rely on any assumptions on the scalar field dynamics; hence it generalizes previous results \cite{Gregory:2018ghc} obtained under the assumption of slow-roll inflation.
The behaviour $\dot{M} \propto M^{2}$ is a feature of the spherical accretion process \cite{Bondi}.   We observe that Eq.~\eqref{Eq:AccretionGeneralMass} is reminiscent of the Bondi accretion formula (see \cite{Celoria:2018mzr,Carr:2010wk} and references therein), although there are two important differences: (i) the rate of change of the mass $M$ on the left-hand side is measured with respect to the null coordinate $v$, as opposed to proper time distant from the hole; (ii) the positive quantity $({\cal L}_l \phi\lvert_{z=0})^2/(1-\mathcal{E})$, which is a local observable at the horizon, replaces the background observable $\rho/c^3$ in the Bondi formula, where $\rho$ and $c$ are the energy density and sound speed of the cosmic fluid, respectively.

\section{Neumann boundary conditions}\label{Sec:Neumann}
\begin{figure}
    \begin{center}
   \includegraphics[width=0.5\columnwidth]{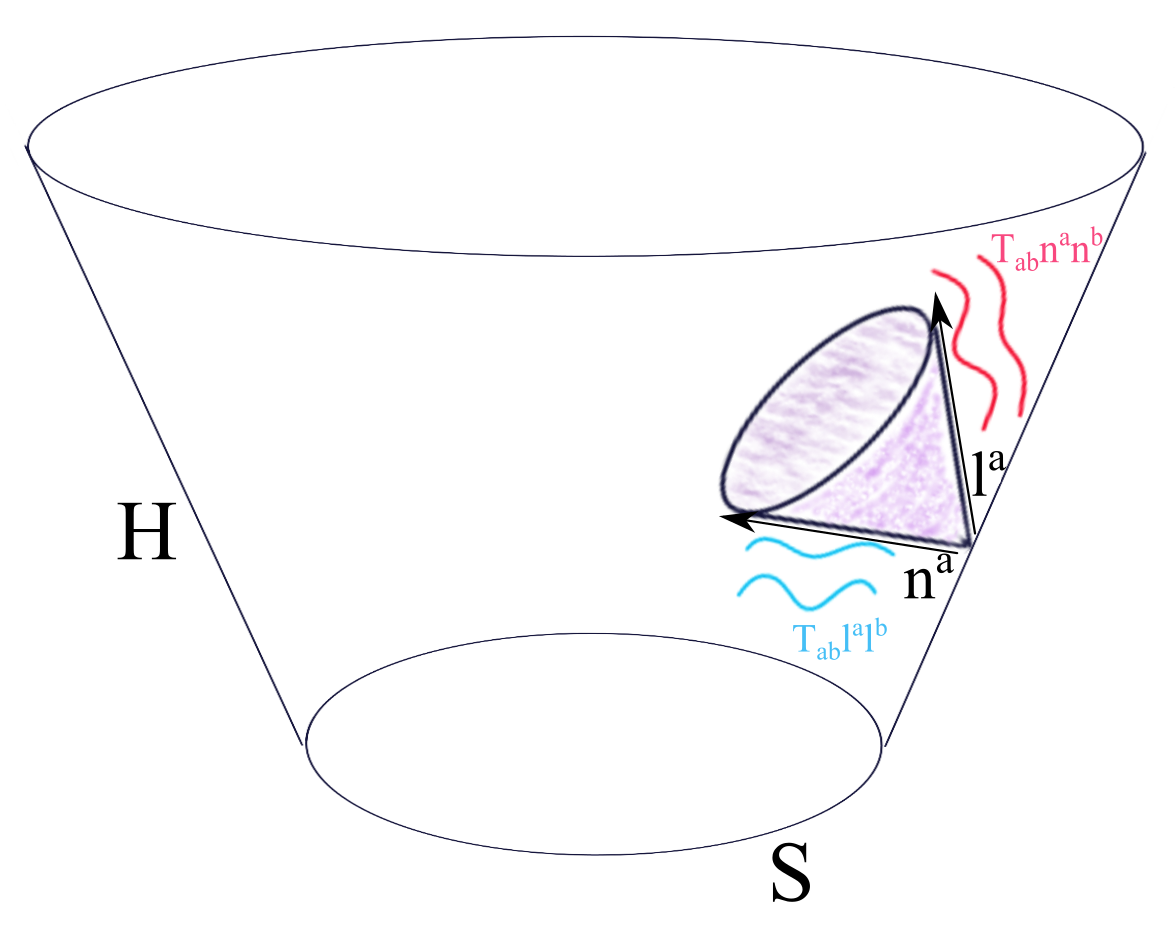}
      \caption{In the figure, $H$ is a three-dimensional spacelike surface tracing the evolution of marginally trapped surfaces $S$ with areal radius $r_H(v)$, see Ref.~\cite{Ashtekar:2004cn}. In a generic configuration the scalar field radiates energy-momentum both in the outgoing and ingoing null directions. In the solution with Neumann boundary conditions at the horizon, the scalar field is drawn inside the black hole merely by the expansion of the horizon; as a result, the flux of energy-momentum is purely ingoing. On the other hand, for a black hole approaching equilibrium the fluxes are in general non-zero in both directions.}
         \label{FigSink}
     \end{center}
\end{figure}

We impose Neumann boundary conditions for the scalar field at the horizon, namely, 
\be\label{Eq:NeumannBC}
(\pa_z \phi)\lvert_{z=0}=0~,
\ee 
which, from Eq.~\eqref{Eq:Expansions}, amounts to $c_1=0$.
As a consequence, the gradient of the scalar field is a null ingoing 1-form at the horizon, $\de \phi\lvert_{z=0}=\dot{\phi}_o \de v$, where the subscript $o$ means that the quantity is evaluated at $z=0$.
As shown in Eq.~\eqref{Eq:StressEnergyNullComponents}, the Neumann boundary condition $c_1=0$ singles out a unique solution, characterized by a purely ingoing energy-momentum flux that is entirely due to the horizon expansion; see Fig.~\ref{FigSink}. The corresponding solutions for the expansion coefficients are reported in Eq.~\eqref{EQ:CoeffSolution} in the Appendix. An accretion law for this solution can be obtained from the more general Eq.~\eqref{Eq:AccretionGeneral} with the substitution ${\cal L}_l \phi\lvert_{z=0}= \dot{\phi}_o$.

The curvature invariants in a neighbourhood of the horizon are all regular for $\dot{r}_H\neq0$, see Eqs.~\eqref{Eq:Horizon-Scalar_System}. However, the solution becomes singular as $\dot{r}_H\to0$ (see Eqs.~\eqref{EQ:CoeffSolution}, \eqref{Eq:Horizon-Scalar_System} in the Appendix), showing that this particular solution is not suitable to approach the static limit. In the next section we show how this problem can be overcome by relaxing the boundary condition~\eqref{Eq:NeumannBC}.

\section{Near-equilibrium solutions}\label{Sec:Adiabatic}
The coordinates $(v,z)$ make the geometry explicitly dependent on the horizon expansion rate $\dot{r}_H$. Furthermore, the parametrization \eqref{Eq:Expansions} allows us to interpret the Einstein equations as an infinite-dimensional dynamical system for the variables $\{a_n, b_n, c_n, \phi_o, r_H\}$ with evolution parameter $v$. This enables us to study the approach to equilibrium by carrying out a systematic expansion of the field equations around a static solution, which represents a fixed point for the system. To this end, we introduce the following perturbative near-equilibrium expansions for the horizon
\be\label{Eq:AdiabaticExpHorizon}
r_H(v)=r_H^{\scr(0)}+r_H^{\scr(1)}(v)+r_H^{\scr(2)}(v)+\dots~,
\ee
and for the dynamical fields
\begin{subequations}\label{Eq:AdiabaticExpFields}
\begin{align}
A(v,z)&= A^{\scr(0)}(z)+A^{\scr(1)}(v,z)+A^{\scr(2)}(v,z)+\dots~,\\
\beta(v,z)&= \beta^{\scr(0)}(v)+\beta^{\scr(1)}(v,z)+\beta^{\scr(2)}(v,z)+\dots~,\\
\phi(v,z)&= \phi^{\scr(0)}+\phi^{\scr(1)}(v,z)+\phi^{\scr(2)}(v,z)+\dots~.\label{Eq:AdiabaticExpFieldsPhi}
\end{align}
\end{subequations}
The quantities $r_H^{\scr(0)}$, $\phi^{\scr(0)}$ are constant, whereas $\beta^{\scr(0)}(v)$ is pure gauge. Similarly, the perturbative expansion of $\phi_o(v)\equiv\phi(v,0)$ is
\be
\phi_o(v)=\phi^{\scr(0)}+\phi^{\scr(1)}_o(v)+\phi^{\scr(2)}_o(v)+\dots ~,
\ee
where $\phi^{\scr(n)}_o(v)=\phi^{\scr(n)}(v,0)$.
We assume the Schwarzschild-de~Sitter geometry for the background
\be\label{Eq:EquilibriumSolution}
A^{\scr(0)}(z)=z+\frac{\kappa}{3} (r_H^{\scr(0)})^2 U( \phi^{\scr(0)})\lf[(1-z) - \frac{1}{(1-z)^2} \rg]~,\quad \beta^{\scr(0)}(v)=0 ~.
\ee
The constant background value of the scalar field must be at a critical point for the potential (this follows from the Klein-Gordon equation~\eqref{EQ:KGequation} to zero-th order in the perturbative expansion)
\be\label{Eq:PotentialCriticalPoint}
\frac{\pa U}{\pa \phi}\bigg\lvert_{ \phi^{\scr(0)}}=0~.
\ee
If the critical point $\phi^{\scr(0)}$ is non-degenerate, the background solution~\eqref{Eq:EquilibriumSolution} corresponds to a one-parameter family of equilibria for the dynamical system, each given by a different value for the constant $r_H^{\scr(0)}$.\footnote{On the other hand, if the potential admits non-isolated critical points there is a two-parameter family of equilibria, parametrized by $r_H^{\scr(0)}$, $\phi^{\scr(0)}$. As an example, one may consider a massless scalar field with constant potential, {\it i.e.}~a cosmological constant.} 
 The existence of degenerate equilibria implies that, in order to correctly account for the near-equilibrium dynamics of the system, we need to include perturbative corrections of order at least two.

We insert the perturbative expansions~\eqref{Eq:AdiabaticExpHorizon}, \eqref{Eq:AdiabaticExpFields} in the field equations~\eqref{Eq:FieldEquations}, to derive the first- and second-order equations for the perturbations. Then, at each order in the near-equilibrium expansion we introduce power-series expansions in $z$ for $A^{\scr(n)}$ and $\beta^{\scr(n)}$ as in Eq.~\eqref{Eq:Expansions}, while for $\phi^{\scr(n)}$ we use the parametrization\footnote{Compared to Eq.~\eqref{Eq:Expansions}, the parametrization \eqref{Eq:PhiAdiabaticExpansion} corresponds to $l^{\scr(n)}_k(v)=(\phi_o(v) c_k(v))^{\scr(n)}$. This is a more convenient choice for the near-equilibrium expansion.}
\be\label{Eq:PhiAdiabaticExpansion}
\phi^{\scr(n)}(v,z) = \phi^{\scr(n)}_o(v) + \sum_{k=1}^{\infty} l^{\scr(n)}_k(v) z^k~.
\ee

The solutions obtained to second order read as
\begin{subequations}\label{Eq:AdiabaticSolutions}
\begin{align}
&A(v,z)= \lf\{1-\kappa\lf[ (r_H^{\scr(0)})^2 \lf(U(\phi^{\scr(0)})+\frac{1}{2}\frac{\pa^2 U}{\pa \phi^2}\bigg\lvert_{\phi^{\scr(0)}}(\phi^{\scr(1)}_o)^2 \rg)  +2 r_H^{\scr(0)} r_H^{\scr(2)}U(\phi^{\scr(0)}) \rg] \rg\} z \\
&\quad -\kappa \lf\{ (r_H^{\scr(0)})^2\lf[ \lf(1- \frac{\kappa}{4} \left(l^{\scr(1)}_1\right)^2\rg)U(\phi^{\scr(0)}) +\frac{1}{2}\left(\phi^{\scr(1)}_o +l^{\scr(1)}_1\right)\phi^{\scr(1)}_o \frac{\pa^2 U}{\pa \phi^2}\bigg\lvert_{\phi^{\scr(0)}}  \rg]  +2r_H^{\scr(0)}r_H^{\scr(2)}U(\phi^{\scr(0)}) +\frac{1}{4}\left(l^{\scr(1)}_1\right)^2  \rg\}z^2 +\dots \nn \\
&\beta(v,z)= \frac{\kappa}{2}\left(l^{\scr(1)}_1\right)^2 z - \frac{\kappa}{4} l^{\scr(1)}_1 \left(l^{\scr(1)}_1 - 4 l^{\scr(1)}_2\right)z^2   +\dots~,\\
&\phi(v,z)= \left(\phi^{\scr(0)}+\phi^{\scr(1)}_o+\phi^{\scr(2)}_o \right)+ \left(l^{\scr(1)}_1+l^{\scr(2)}_1\right) z + \left(l^{\scr(1)}_2+l^{\scr(2)}_2 \right) z^2 +\dots~.
\end{align}
\end{subequations}
The coefficients $l^{\scr(1)}_1$ and $l^{\scr(1)}_2$ satisfy the following differential equations
\begin{subequations}\label{Eq:Solutions_lncoeffs}
\begin{align}
\dot{l}^{\scr(1)}_1 &= -\dot{\phi}^{\scr(1)}_o + \frac{r_H^{\scr(0)}}{2}\phi^{\scr(1)}_o \frac{\pa^2 U}{\pa \phi^2}\bigg\lvert_{\phi^{\scr(0)}} -
\frac{l^{\scr(1)}_1}{2 r_H^{\scr(0)}}\Big(1-\kappa\,(r_H^{\scr(0)})^2 U(\phi^{\scr(0)}) \Big)~,\label{Eq:Solutions_lncoeffsA}\\
\dot{l}^{\scr(1)}_2 &= \frac{r_H^{\scr(0)}}{4} \phi^{\scr(1)}_o \frac{\pa^2 U}{\pa \phi^2}\bigg\lvert_{\phi^{\scr(0)}}  + \frac{l^{\scr(1)}_1}{4r_H^{\scr(0)}}  \lf[ 3 - (r_H^{\scr(0)})^2 \lf( \kappa\, U(\phi^{\scr(0)}) - \frac{\pa^2 U}{\pa \phi^2}\bigg\lvert_{\phi^{\scr(0)}} \rg)\rg]- \frac{l^{\scr(1)}_2}{r_H^{\scr(0)}}\Big(1- \kappa (r_H^{\scr(0)})^2 U(\phi^{\scr(0)})\Big)~.
\end{align}
\end{subequations}
The first non-trivial correction to the horizon arises at second order in the perturbative expansion,\footnote{More specifically, to first order we have $\dot{r}_H^{\scr(1)}=0$; hence, the constant value of $r_H^{\scr(1)}$ can be reabsorbed into $r_H^{\scr(0)}$.} which yields the accretion law
\be\label{Eq:AdiabaticAccretionLaw}
\dot{r}_H^{\scr(2)}=\frac{\kappa (r_H^{\scr(0)})^2 }{1- {\cal E}^{\scr(0)}}(\dot{\phi}^{\scr(1)}_o)^2~,
\ee
where ${\cal E}^{\scr(0)}=\kappa\,(r_H^{\scr(0)})^2 U(\phi^{\scr(0)})$. 
We remark that the approximate solutions for $A$ and $\beta$ \eqref{Eq:AdiabaticSolutions}, as well as Eqs.~\eqref{Eq:Solutions_lncoeffs}, \eqref{Eq:AdiabaticAccretionLaw}, only depend on the zeroth- and first-order solutions for $\phi$; for this reason the coefficients $\phi^{\scr(2)}_o$ and $l^{\scr(2)}_n$ have no influence on the geometry to this order of approximation.

In contrast to the Neumann solution studied in Section~\ref{Sec:Neumann}, the coefficient $c_1\simeq l^{\scr(1)}_1/\phi^{\scr(0)}$ is not identically zero for a generic near-equilibrium solution; its value only relaxes to zero in the approach to equilibrium. Using the results in Section~\ref{Sec:Geo}, we observe that such solutions have non-zero fluxes both in the ingoing and the outgoing null directions: to first order in perturbation theory we have
\be
T_{ab}n^a n^b \Big\lvert_{z=0}\simeq  \frac{(l^{\scr(1)}_1)^2}{(r_H^{\scr(0)})^2}~,~~T_{ab}l^a l^b\Big\lvert_{z=0}\simeq (\dot{\phi}^{\scr(1)}_o)^2 ~.
\ee
Indeed, it follows from Eq.~\eqref{Eq:Solutions_lncoeffsA} that the fluxes are both zero only in the static limit $\dot{\phi}^{\scr(1)}_o, l^{\scr(1)}_1\to 0$.
In particular, we observe that if $l^{\scr(1)}_1= 0$ identically then Eq.~\eqref{Eq:Solutions_lncoeffsA} implies that the scalar field must climb up the potential, which is incompatible with the approach to equilibrium. This result explains why the equilibrium cannot be approached if the flux in the outgoing null direction $T_{ab}n^a n^b\lvert_{z=0}$ is identically zero (as for the solution studied in Sec.~\ref{Sec:Neumann}). In the remainder of this section we determine conditions for attaining equilibrium dynamically. 

\subsection{Approaching equilibrium}
In order to fully determine the near-equilibrium solutions using the results in Section~\ref{Sec:Adiabatic}, we need to assign the functional form of $\phi^{\scr(1)}_o(v)$. By analogy with the structure of Eq.~\eqref{Eq:Solutions_lncoeffsA}, we shall assume a simple model where $\dot{\phi}^{\scr(1)}_o(v)$ depends linearly on $\phi^{\scr(1)}_o$ and $l^{\scr(1)}_1$:  $r_H^{\scr(0)} \dot{\phi}^{\scr(1)}_o = -\gamma\, \phi^{\scr(1)}_o + \xi\, l^{\scr(1)}_1$. In this way, using Eq.~\eqref{Eq:Solutions_lncoeffsA}, the dynamics of $\phi^{\scr(1)}_o$ and $l^{\scr(1)}_1$ are described by the following autonomous dynamical system
\begin{subequations}\label{Eq:DynSys}
\begin{align}
r_H^{\scr(0)} \dot{\phi}^{\scr(1)}_o &= -\gamma\, \phi^{\scr(1)}_o + \xi\, l^{\scr(1)}_1~,\\
r_H^{\scr(0)} \dot{l}^{\scr(1)}_1 &=  \lf(\gamma +\frac{(r_H^{\scr(0)})^2}{2}\frac{\pa^2 U}{\pa \phi^2}\bigg\lvert_{\phi^{\scr(0)}} \rg) \phi^{\scr(1)}_o -\frac{1}{2} \Big(1-{\cal E}^{\scr(0)}+2\xi \Big)  l^{\scr(1)}_1 ~.
\end{align}
\end{subequations}
Here $\gamma$ and $\xi$ are assumed to be constant dimensionless parameters for simplicity, although generalisations are possible; the actual values of these parameters need to be determined by suitably matching to an outer solution describing the spacetime far from the black hole. 

Introducing a dimensionless ``time'' parameter $T=v/r_H^{\scr(0)}$ and the matrix notation
\be\label{Eq:DynSysMatrix}
{\bf X}=\begin{pmatrix} \phi^{\scr(1)}_o \\ l ^{\scr(1)}_1 \end{pmatrix} ~, \quad {\bf A}= \begin{pmatrix} -\gamma & \xi \\ \gamma +\frac{(r_H^{\scr(0)})^2}{2}\frac{\pa^2 U}{\pa \phi^2}\Big\lvert_{\phi^{\scr(0)}}  & -\frac{1}{2}(1-{\cal E}^{\scr(0)}+2\xi)  \end{pmatrix} ~,
\ee
we can rewrite Eq.~\eqref{Eq:DynSys} as
\be
\frac{\de{\bf X}}{\de T}= {\bf A} \cdot {\bf X}~.
\ee

The dynamical system \eqref{Eq:DynSys} can be studied with standard methods. The unperturbed static solution with $\phi^{\scr(1)}_o=0=l^{\scr(1)}_1$  is an attractive fixed point provided that the eigenvalues of the coefficient matrix in \eqref{Eq:DynSysMatrix} are both negative, that is 
\begin{subequations}
\begin{align}
&2(\gamma+\xi)+1-{\cal E}^{\scr(0)} >0 ~,\\
&0<\gamma \lf(1-{\cal E}^{\scr(0)}\rg)-\xi (r_H^{\scr(0)})^2 \frac{\pa^2 U}{\pa \phi^2}\bigg\lvert_{\phi^{\scr(0)}} \leq 
\frac{1}{8} \Big[2(\gamma+\xi)+1-{\cal E}^{\scr(0)}\Big]^2 ~.
\end{align}
\end{subequations}
Under this assumption, a scalar field initially perturbed away from its equilibrium $\phi^{\scr(0)}$ (assumed as non-degenerate) will eventually settle down into its equilibrium configuration. Meanwhile, the horizon evolves from its initial value $r_H^{\scr(0)}$ according to Eq.~\eqref{Eq:AdiabaticAccretionLaw}, reaching a final value $r_H^{\,\it f} > r_H^{\scr(0)}$. Thus, the black hole evolution represents a transition between two Schwarzschild-de~Sitter solutions, with the same cosmological constant $\Lambda_{\rm eff}=\kappa\, U(\phi^{\scr(0)})$ but different masses.

Let us examine the solutions of the system \eqref{Eq:DynSys} more in detail, assuming $\xi\neq0$ for definiteness. These are given by
\be
\phi_o^{\scr(1)}(v)=p\, e^{\lambda_1 v/r_H^{\scr(0)}}+q\, e^{\lambda_2 v/r_H^{\scr(0)}} ~, \quad l^{\scr(1)}_1(v)=\xi^{-1}p(\gamma+\lambda_1)\, e^{\lambda_1 v/r_H^{\scr(0)}}+\xi^{-1}q(\gamma+\lambda_2)\, e^{\lambda_2 v/r_H^{\scr(0)}}~,
\ee
where $p$ and $q$ are (real) constant coefficients, while $\lambda_{1}$ and $\lambda_{2}$ are solutions of the eigenvalue equation for the matrix ${\bf A}$
\be
2\lambda ^2+\lf(2\gamma+2 \xi+1-{\cal E}^{\scr(0)}\rg)  \lambda +\gamma\lf(1 -{\cal E}^{\scr(0)} \rg)   -\xi  (r_H^{\scr(0)})^2 \frac{\pa^2 U}{\pa \phi^2}\bigg\lvert_{\phi^{\scr(0)}}=0~.
\ee
We consider the case where both eigenvalues are negative real numbers, so that $\phi^{\scr(1)}_o=l^{\scr(1)}_1=0$ is an attractive fixed point. We also label the eigenvalues so that $\lambda_2 < \lambda_1$.\footnote{We do not consider the case of degenerate eigenvalues, which requires fine-tuning.} Thus, we have in the large-$v$ limit
\be
\phi_o^{\scr(1)}(v)\sim p\, e^{\lambda_1 v/r_H^{\scr(0)}} ~,\quad l^{\scr(1)}_1(v)\sim \xi^{-1}p(\gamma+\lambda_1)\, e^{\lambda_1 v/r_H^{\scr(0)}}~.
\ee
Substituting in the accretion law \eqref{Eq:AdiabaticAccretionLaw} we obtain 
\be
\dot{r}_H^{\scr(2)}(v)\sim\frac{\kappa  p^2 \lambda_1^2  e^{2 \lambda_1 v/r_H^{\scr(0)}}}{1- {\cal E}^{\scr(0)}}\implies r_H^{\scr(2)}(v)\sim\Delta r_H+ \frac{\kappa  p^2 \lambda_1  }{2(1- {\cal E}^{\scr(0)})}r_H^{\scr(0)}\, e^{2 \lambda_1 v/r_H^{\scr(0)}}~,
\ee
where we introduced the integration constant $\Delta r_H=r_H^{\, \it f}-r_H^{\scr(0)}$. Thus, the evolution of the apparent horizon has the following asymptotics
\be
r_H(v)\sim r_H^{\, \it f}+ \frac{\kappa  p^2 \lambda_1  }{2(1- {\cal E}^{\scr(0)})}r_H^{\scr(0)}\, e^{2 \lambda_1 v/r_H^{\scr(0)}}~.
\ee
Combining the above results we get the following scaling laws describing the approach to equilibrium
\be\label{Eq:Scaling}
\lf\lvert\frac{r_H(v)-r_H^{\it f}}{r_H^{\scr(0)}}\rg\lvert\sim \kappa(\phi_o^{\scr(1)}(v))^2~,\quad l^{\scr(1)}_1(v)\sim\phi_o^{\scr(1)}(v)~.
\ee
Except for numerical pre-factors, these scaling relations hold regardless of the specific functional form of the scalar potential and numerical values of parameters such as $\gamma$ and $\xi$.

%%%%%%%%%%%%%%%%%
%SECTION 3
%%%%%%%%%%%%%%%

\section{First law of black hole dynamics}\label{Sec:Thermodynamics}
As shown in previous sections, the evolution of the trapping horizon is fully characterised by two dynamical variables: $r_{H}(v)$ and $\phi_{o}(v)$. Thus, all thermodynamical properties of a black hole in the presence of a scalar field and in the proximity of the horizon can be expressed in terms of these two quantities. In this section we derive the first law and the Smarr formula, which hold for any solutions of the field equations.

The surface gravity $g$ for an evolving horizon can be computed using the geometric definition given in Ref.~\cite{Hayward:2008jq}
\be
g(v)\equiv \lf. \frac{1}{2}\star d \star d r \rg|_{r=r_H(v)} =\frac{1}{2}\left( \pa_r A +A \pa_r \beta \right) \lvert_{r=r_H(v)} ~,
\ee
where $\star$ is the Hodge star operator in the two-dimensional timelike surface orthogonal to the two spheres. Using the approximated solution derived in Appendix and evaluating the result at $z=0$, this gives
\be
g=\frac{1- {\cal E}}{2 r_H}=\frac{1- {\cal E}}{4 G M} ~.
\ee
This result can be rearranged as a Smarr formula
\be\label{Eq:Smarr1}
M=\frac{g}{4\pi G}{\cal A}+3 U(\phi_o) \cal{V} ~.
\ee
The mass is a homogeneous function of degree $1/2$ of ${\cal A}\equiv 4\pi r_H^2$ and ${\cal{V}}^{2/3} = \left[(4/3) \pi r^{3}_{H}\right]^{2/3}$. Consistently, the first law reads as
\be\label{Eq:FirstLaw2}
 \delta M= \frac{g}{8\pi G}\delta{\cal A}+ U(\phi_o)\, \delta\cal{V}~,
\ee
in agreement with the first law of black hole dynamics as given in Ref.~\cite{Hayward:1997jp} (see Eq.~(7.1) therein). Our result, however, contrasts with the first law derived in \cite{Gregory:2018ghc} (see Eq.~(95) therein).

Equation~\eqref{Eq:FirstLaw2} also suggests an intriguing analogy with the dynamics of elastic membranes: the scalar potential can be interpreted as a tension (negative pressure\footnote{This interpretation is also consistent with $T^r_{\; r}=  -U(\phi_o)+\mathcal{O}(z)$.}) $\tau\equiv U(\phi_o)$. It would be interesting to further explore this analogy within the context of the {\it membrane paradigm} \cite{1982:damour,1986:thorne}.
The first law of black hole dynamics \eqref{Eq:FirstLaw2} considers the scalar field surrounding the black hole as an external source. In other words, the black hole is considered immersed in a thermal bath given by the scalar field. The second term on the right-hand side of the first law \eqref{Eq:FirstLaw2} should be interpreted as the work done by the scalar field along the horizon. Finally, we would like to emphasise that this thermodynamical interpretation proposed in Eq.~\eqref{Eq:FirstLaw2} reinforces the understanding of the dynamical black hole with scalar field in the close proximity of the horizon as a sequence of Schwarzschild-de Sitter black holes with a (locally defined) effective cosmological constant, proportional to the scalar potential $U(\phi_o)$.

\section{Discussion}\label{Sec:Discussion}
We obtained for the first time approximate analytical solutions for an evolving black hole in the presence of a self-interacting scalar field. The solutions obtained admit a power-series expansion in $z$, a radial coordinate measuring the displacement from the black hole apparent horizon. We derived an exact accretion law~\eqref{Eq:AccretionGeneralMass}, which represents a fully relativistic generalization of the Bondi accretion formula.
In the special case of Neumann boundary conditions at the horizon, the solution takes the simplest form and its expansion coefficients are given explicitly in the Appendix, along with the corresponding curvature invariants. In this solution the scalar field falls inside the black hole without emitting an outward flux of energy-momentum. However, we have shown that this solution cannot approach the static limit.

We also obtained near-equilibrium solutions in Section~\ref{Sec:Adiabatic}, obtained by solving the field equations perturbatively around the static Schwarzschild-de~Sitter solution.
Our choice of coordinates $(v,z)$ is particularly convenient for studying the approach to equilibrium, since the metric in these coordinates explicitly depends on the expansion rate $\dot{r}_H$. We show that in this regime the evolution of the system can be described as a dynamical system; then, we explicitly obtain the solutions in a simple model, showing that the approach to equilibrium is characterized by universal scaling relations. 
Future work will be devoted to the matching of the solutions here obtained in a neighbourhood of the horizon to the region far from the black hole, where the scalar field follows its cosmological evolution.
This will enable us to study the accretion of black holes during inflation without simplifying assumptions, such as slow roll. In the case of asymptotically flat solutions, it would be interesting to study the evolution of the apparent horizon during gravitational collapse in the super-critical regime ({\it i.e.}, above the threshold for black-hole formation).
Both in the asymptotically flat and in the cosmological cases, our analytical methods will offer a useful complement to numerical relativity simulations. The generalization to the axisymmetric case will be studied in a future work.
The solution techniques here illustrated have much broader applicability: they can be applied to different matter fields coupled to gravity (including, \emph{e.g.}, hydrodynamic matter and gauge fields) as well as modified gravity theories. For instance, in the case of a complex scalar field one can follow similar steps, which lead to the accretion law $\dot{M}=16 \pi  G^2/(1-8\pi G r_H^2 U(\phi^{*}\phi))  M^2 |{\cal L}_l \phi|^2$ instead of \eqref{Eq:AccretionGeneralMass}.
Our methods can also be used to trace the exact evolution of the black hole apparent horizon in scalar-tensor theories, extending previous works, \emph{e.g.}, \cite{Jacobson:1999vr}.
We expect that the evaporation process due to Hawking radiation \cite{Parentani:1994ij} or dark energy \cite{Babichev:2004yx} can also be described using similar methods, by including appropriate couplings to sources that violate the energy conditions. 

\begin{acknowledgments}
We are grateful to Ivan Booth, Eric Gourgoulhon, Viqar Husain, Jos{\'e} M.M.~Senovilla, and Ra{\"u}l Vera for helpful comments on an earlier draft of this paper. MdC also thanks Ra{\"u}l Vera and Jos{\'e} M.M.~Senovilla for many stimulating discussions.
The work of MdC is supported under grants No. FIS2017-85076-P (Spanish MINECO/AEI/FEDER, UE) and No. IT956-16 (Basque Government).
The work of RO is supported by the R{\'e}gion {\^I}Žle-de-France, within the DIM ACAV$^{+}$ SYMONGRAV project. This work also received funding by the European Structural and Investment Funds (ESIF) and the Czech Ministry of Education, Youth and
Sports (MSMT), Project CoGraDS - CZ.02.1.01/0.0/0.0/15003/0000437.
\end{acknowledgments}

\appendix
\renewcommand{\theequation}{A.\arabic{equation}}

\section{Coefficients for the Neumann solution}\label{Sec:Appendix}

Plugging the expansions \eqref{Eq:Expansions} into Eqs.~\eqref{Eq:FieldEquations}, \eqref{EQ:KGequation}, we obtain equations for the unknown expansion coefficients.
As an explicit example, the coefficients for the solution with Neumann boundary conditions are, for the first three orders in~$z$, given by
\begin{subequations}\label{EQ:CoeffSolution}
\begin{align}
a_1 &= 1-{\cal E} >0 ~,  &a_2 &= a_1-1 ~,  &a_3 &=-\frac{1}{12} \lf[\frac{4 (3-4 {\cal E}) {\cal E}}{1-{\cal E}}+\frac{2 r_H (1+{\cal E})\dot{{\cal E}}}{{\cal K}}-\frac{r_H^2 (1-{\cal E})\dot{{\cal E}}^2}{{\cal K}^2}\rg]~, \\ 
b_{1} &=0~,  &b_2 &= 0 ~,   &b_3&= \frac{\left((1-{\cal E})r_H  \dot{{\cal E}}-2 {\cal K}\right)^2}{24{\cal K}^3}~,\\
c_1&= 0 ~, & c_2 &= \frac{r_H }{2 {\cal K} }\frac{\dot{\phi}_o}{\phi_o}-\frac{r_H^2 (1-{\cal E}) \dot{{\cal E}} \dot{\phi}_o}{4{\cal K}^2 \phi_o} ~,
\end{align}
and
\begin{align}
 c_3=\frac{r_H^2}{12 {\cal K}^3\phi_o} \left\{\dot{\phi}_o r_H^{-1} \left[
   r_H^2(1-{\cal E}) \left(\dot{{\cal E}}^2-(1-{\cal E})\ddot{{\cal E}}\right)+2 {\cal K}\Big(1+{\cal K}+{\cal E} ({\cal E}+{\cal K}-2)\Big)+\rg.\rg.\nonumber\\
  \lf.\lf. -r_H(1-{\cal E})\left({\cal K}+(1-{\cal E})^2\right) \dot{{\cal E}}\right] -(1-{\cal E}) \left(2{\cal K}- 3 r_H (1-{\cal E}) \dot{{\cal E}}\right)\ddot{\phi}_o \rg\}~,
\end{align}
\end{subequations}
We recall the definitions $\mathcal{E}\equiv\kappa\, r_H^2 U(\phi_o)$ and ${\cal K}\equiv \kappa\,  r_H^2 (\dot{\phi}_o)^2$.
Higher-order coefficients can be computed order by order. Note that in the $\dot{r}_H\to0$ limit some coefficients diverge, signalling that the static limit is singular; for instance, even in the case of a constant potential the coefficient $b_3$ diverges as $\sim \dot{\phi}_o^{-4}$.
The curvature invariants in a neighbourhood of the horizon are all regular for $\dot{r}_H\neq 0$; their expansions read as
\begin{subequations}\label{Eq:Horizon-Scalar_System}
\begin{align}
R&=\frac{1}{r_H^2}\lf\{4{\cal E} +\left[2-r_H (1-{\cal E})\frac{\dot{{\cal E}}}{{\cal K}}\right]z+ \mathcal{O}(z^2) \rg\}~,\\
R_{ab}R^{ab}&=\frac{1}{r_H^4}\lf\{4{\cal E}^2 + 2 {\cal E} \left[2- r_H (1-{\cal E}) \frac{\dot{{\cal E}}}{{\cal K}}\right] z+ \mathcal{O}(z^2) \right\} ~,\\
R_{abcd}R^{abcd} &=\frac{4}{r_H^4} \Bigg\{ 3-(2-{\cal E}) {\cal E} - \lf[2 ({\cal E}-4) ({\cal E}-2) +r_H (1-{\cal E}) \frac{\dot{{\cal E}}}{{\cal K}}\rg] z+ \mathcal{O}(z^2) \Bigg\} ~, \\ 
C_{abcd}C^{abcd} &=\frac{4}{3 r_H^4} (3-{\cal E}) \Bigg\{ (3-{\cal E})-\lf[16 -6 {\cal E}+r_H (1-{\cal E}) \frac{\dot{{\cal E}}}{{\cal K}}\rg] z + \mathcal{O}(z^2)\Bigg\}~.
\end{align}
\end{subequations}
Here we recall that $\mathcal{E}\equiv\kappa\,r_H^2 U(\phi_o) $ and ${\cal K}\equiv \kappa\,  r_H^2 (\dot{\phi}_o)^2$.

\bibliography{NHrefs}

\end{document}